\documentclass[conference]{IEEEtran}
\IEEEoverridecommandlockouts
\usepackage[noadjust]{cite}
\usepackage{amsmath,amssymb,amsfonts}
\usepackage{algorithmic}
\usepackage{graphicx}
\usepackage{textcomp}
\usepackage{xcolor}
\usepackage{float}
\usepackage{enumitem}
\usepackage{booktabs} 
\usepackage{csquotes}
\usepackage{multirow}
\usepackage{soul}
\usepackage{balance}

\def\BibTeX{{\rm B\kern-.05em{\sc i\kern-.025em b}\kern-.08em
    T\kern-.1667em\lower.7ex\hbox{E}\kern-.125emX}}

\title{Investigating the Essential of Meaningful Automated Formative Feedback for Programming Assignments}

\begin{document}

\author{\IEEEauthorblockN{Qiang Hao\IEEEauthorrefmark{1},
Jack P Wilson\IEEEauthorrefmark{2}, Camille Ottaway\IEEEauthorrefmark{3}, Naitra Iriumi\IEEEauthorrefmark{4}, Kai Arakawa\IEEEauthorrefmark{5} and David H Smith IV\IEEEauthorrefmark{6}
}
\IEEEauthorblockA{Western Washington University\\
Bellingham, WA, USA\\
Email: \IEEEauthorrefmark{1}qiang.hao@wwu.edu,
\IEEEauthorrefmark{2}wilso313@wwu.edu,
\IEEEauthorrefmark{3}ottawac@wwu.edu,\\
\IEEEauthorrefmark{4}iriumin@wwu.edu,
\IEEEauthorrefmark{5}hicksk5@wwu.edu,
\IEEEauthorrefmark{6}smithd77@wwu.edu
}}

\IEEEoverridecommandlockouts
\IEEEpubid{\makebox[\columnwidth]{978-1-7281-0810-0/19/\$31.00~\copyright2019 IEEE \hfill} \hspace{\columnsep}\makebox[\columnwidth]{ }}

\maketitle

\IEEEpubidadjcol

\begin{abstract}
This study investigated the essential of meaningful automated feedback for programming assignments. Three different types of feedback were tested, including (a) \textit{What's wrong} - what test cases were testing and which failed, (b) \textit{Gap} - comparisons between expected and actual outputs, and (c) \textit{Hint} - hints on how to fix problems if test cases failed. 46 students taking a CS2 participated in this study. They were divided into three groups, and the feedback configurations for each group were different: (1) Group One - \textit{What's wrong}, (2) Group Two - \textit{What's wrong} + \textit{Gap}, (3) Group Three - \textit{What's wrong} + \textit{Gap} + \textit{Hint}. This study found that simply knowing what failed did not help students sufficiently, and might stimulate system gaming behavior. Hints were not found to be impactful on student performance or their usage of automated feedback. Based on the findings, this study provides practical guidance on the design of automated feedback.
\end{abstract}


\section{Introduction}

Student interest in computer science (CS) has increased substantially over the last decade. In the U.S., undergraduate CS enrollment has doubled since 2011, and class sizes of programming courses offered in colleges have more than tripled \cite{Camp2017}. CS courses nowadays are characterized by large enrollments and low instructor-to-student ratios, especially for the entry-level CS courses, such as CS1, CS2 or data structures \cite{Camp2017, sax2017examining}. The challenges in assessing programming assignments of a large number of students make it difficult for students to get feedback in time. When students work on either individual or group programming assignments, they may meet challenges they can not overcome. If feedback can be provided at those moments when it is needed the most, the learning efficacy can be significantly enhanced.

Prior studies addressing the feedback challenge originated from automated grading of programming assignments. As class sizes grew rapidly, it was natural to ensure that programming assignments were assessed in a timely manner \cite{CHEANG2003121, Singh2013}. The investigation on auto-grading made significant contributions to computing education research, but also cast a perspective of summative feedback on the efforts to automate feedback --- feedback should be provided along the assessment results \cite{Brinko1993, GIELEN2010304}. A popular concern was that formative feedback, the feedback provided during the learning process, may lead to students gaming the system \cite{Hattie2007, Chen2004, ihantola2010}. Many tested programming assignment systems provide no formative feedback to students or limit the allowed number of submissions \cite{ihantola2010, safei2014, akccapinar2015}. As a result, there is a gap in our understanding on how students utilize automated formative feedback and whether that leads to better learning efficacy.

To fill this gap, this study investigated the essential components of effective automated formative feedback through a controlled experiment. The results of this study provide empirical evidence on the efficacy of automated formative feedback of different configurations, and contribute to the understanding of how CS students utilize it for just-in-time learning. 

\section{Related Works}

\subsection{Automated Grading and Feedback}
Programming assignments are difficult to assess and provide feedback in a timely manner for many reasons, including multiple possible approaches to problem solving, necessity to test against many cases to reach sufficient test coverage, and different individual coding habits and styles \cite{CHEANG2003121, sztipanovits2008, Guerreiro2006}. As student enrollment grows, the first challenge to address was the assessment. As a result, automated grading has been investigated extensively.

Studies before 2010 on this topic tended to focus on automated grading system development and testing \cite{Guerreiro2006, malmi2005, ihantola2010}. Systems developed in this period of time require instructors to provide representative test cases and manually tune feedback to work effectively. Web-CAT and Autolab are two representative examples \cite{edwards2008web, haldeman2018providing}. More importantly, early systems and studies had great concerns over the possibility that students may game the system, so such systems typically expected the submission of a fully completed program before providing feedback, limit the number of submissions, and limit the completeness of the feedback \cite{Guerreiro2006, Chen2004, ihantola2010}.

A focus shift from automated grading to automated feedback was witnessed in the most recent decade. Specifically, the focus was on  feedback generation through data-driven approaches \cite{gao2016, head2017}. Massive Open Online Programming Courses provided large datasets of programming assignments, which is critical to make such approaches possible. The efforts typically aim at providing student suggestions on repairing their program through measuring the distance between their program and the most similar working program \cite{gulwani2018, parihar2017, wang2018}. Such efforts are still in their early stage for two reasons. First, these approaches were rarely tested in authentic environments. Second, how these approaches can be effectively applied to a significantly smaller dataset (i.e., a face-to-face CS1 in a large university) is still unknown.

\subsection{Effective Feedback from Educational Perspectives}

In general, feedback can be categorized into two types: formative and summative. Summative feedback is provided when assessment results are released, whereas formative feedback is provided during the learning process. Formative feedback has been found constantly more effective than summative feedback in helping students learn in educational studies across different disciplines, because summative feedback serves more as a justification of the assessment results in student eyes \cite{Brinko1993, GIELEN2010304, Chen2004}. However, this important perspective was not well taken by studies on automated feedback. Early studies found that students tended to abuse multi-leveled hints where the bottom-level hint revealed direct answers \cite{roll2011improving, baker2008students, aleven2016help}. Although hints can only be considered as one type of feedback, many studies used the two terms interchangeably \cite{eagle2013evaluation, Guerreiro2006, Chen2004, roll2011improving}. This may contribute to the lack of investigations on how CS students actually use automated feedback.

\begin{figure}[H]
\centering
\includegraphics[width=90mm]{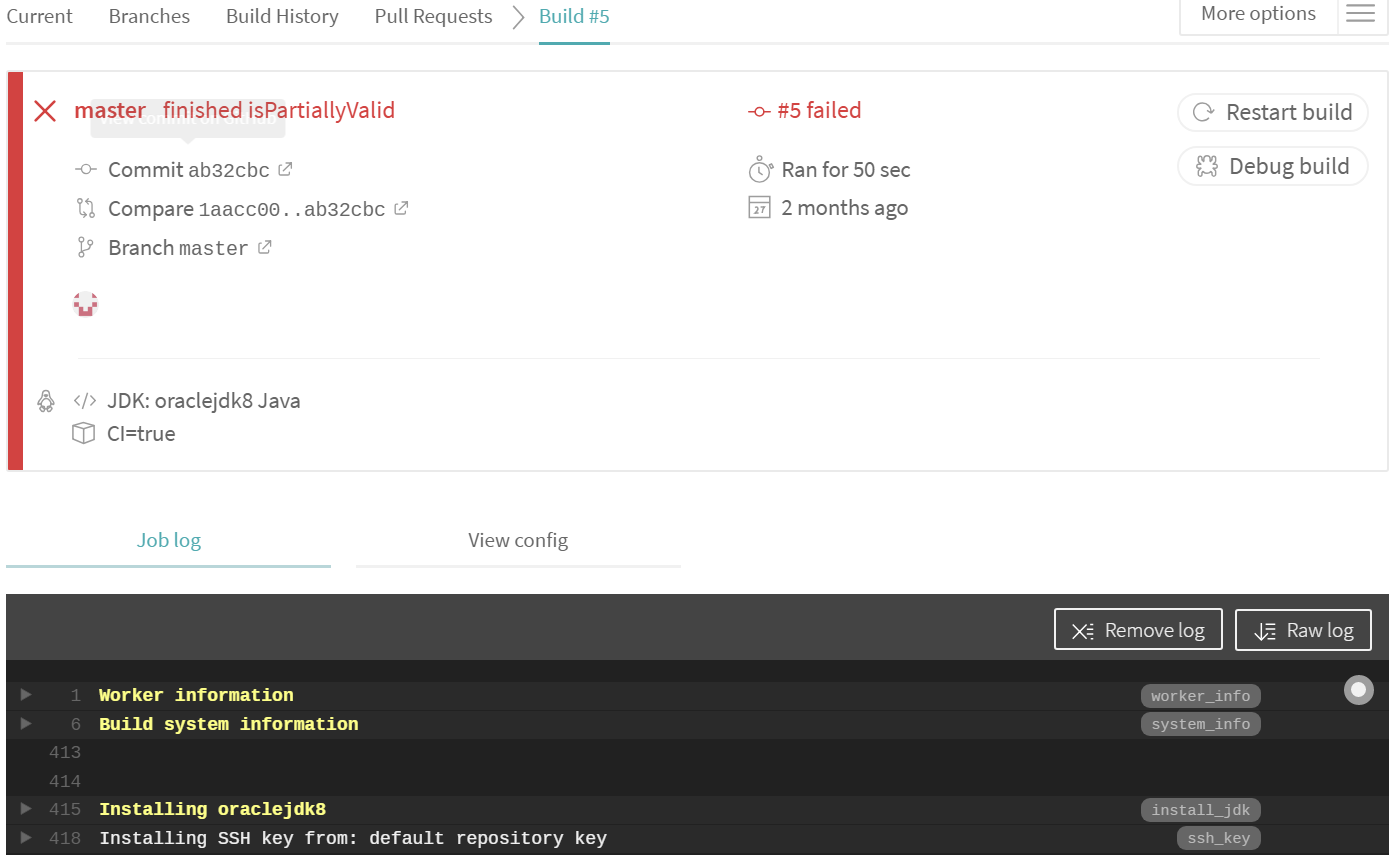}
\caption{A partial screenshot of what a student sees on Travis-CI}
\label{fig:sys}
\end{figure}

\begin{figure}[b]
\centering
\includegraphics[width=90mm]{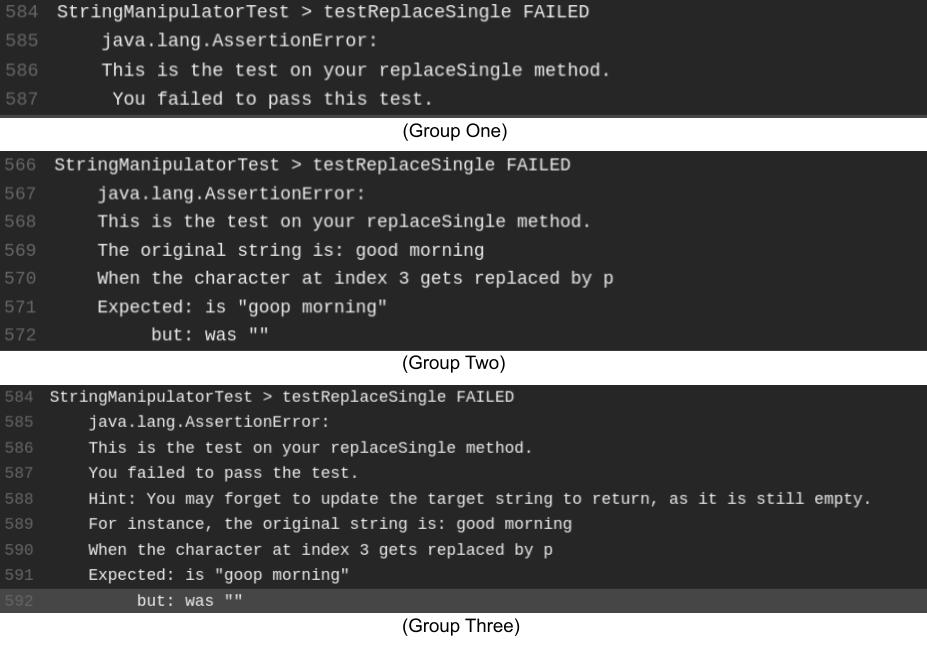}
\caption{An example of different feedback configurations for a method}
\label{fig:comparison}
\end{figure}

\section{Research Design}
\subsection{Research Questions}

Two research questions to guide the research design, including:

\begin{enumerate}
  \item What feedback component is more essential than others to student learning?
  \item How do automated feedback configurations affect student usage behavior?
\end{enumerate}

\subsection{System Implementation and Feedback Design}

The automated feedback service we implemented for this study was through the existing infrastructures, including GitHub, GitHub Classroom, Travis-CI and Gradle. Travis-CI is a distributed continuous integration service for building and testing software projects hosted at GitHub \cite{travis}. Gradle is an open-source build automation system for Java programming \cite{gradle}. Our settings allow students to commit \& push to GitHub as many times as they want before due dates. Every commit \& push will trigger the execution of the submitted code and prepared testing code, and the generated formative feedback will be presented on Travis-CI (https://travis-ci.com; see Figure \ref{fig:sys}). Students can always get formative feedback regardless of if they have fully finished the program or not. In addition, no explicit submission actions are required beyond regular commits \& pushes.

The key of this study is the design of feedback. Synthesizing the literature on feedback, we classified common feedback into three types in the context of computing education:

\begin{enumerate}
  \item \textit{What's wrong}: Per test case, what it is testing and whether it is a pass or failure \cite{narciss2014exploring}
  \item \textit{Gap}: Per test case, what the expected output is, and what the actual output is \cite{metcalfe2007principles}
  \item \textit{Hint}: Per test case, how you may fix the issue if the test case fails \cite{roll2011improving, eagle2013evaluation}
\end{enumerate}

All types of feedback were implemented per test case. To effectively implement \textit{Hint}, we adopted the approach used by Parihar et al. \cite{parihar2017}. We collected student submissions of the same programming assignments over the last three quarters, summarized the common mistakes and problems, and designed adaptive hints for the top five common mistakes per test case.

\subsection{Experiment Design}

This study was conducted in a large university in the North American Pacific Northwest. 46 students taking a CS2 participated in this study. The course was composed of both lectures and lab sessions. Students were expected to complete three complex individual programming assignments during the lab sessions. Each assignment required about 200-300 lines of code to completely solve the given problem. Students were randomly and evenly divided into three different lab sessions. Each student only attended one lab session throughout the whole semester. Therefore, each lab session was treated as a group. The automated feedback each group received on their programming assignments were configured different (see Figure \ref{fig:comparison}): 

\begin{enumerate}
  \item Group One: \textit{What's wrong}
  \item Group Two: \textit{What's wrong} + \textit{Gap}
  \item Group Three: \textit{What's wrong} + \textit{Gap} + \textit{Hint}
\end{enumerate}

All students learned how to use Git before taking the experimental course, and they were given detailed instruction on how to utilize automated formative feedback in the beginning of the course. All student coding behaviors captured by Git and Travis were tracked. Students were asked to indicate the frequency of checking feedback on Travis-CI by the end of the course. Their performance on programming assignments were also recorded.

\section{Results}

\subsection{What feedback component is more essential than others to student learning?}

To answer the question "\textit{What feedback component is more essential than others to student learning?}", we examined and compared student programming assignment performance across the three groups. The performance of 46 students from three groups was summarized in Table 1.

\begin{table}[H]
\caption{Student average performance per group per programming assignment}
\begin{tabular}{lllll}
\hline
\multirow{2}{*}{Group} & \multirow{2}{*}{\begin{tabular}[c]{@{}l@{}}Student \\ numbers\end{tabular}} & \multicolumn{3}{c}{Average Performance}            \\ \cline{3-5} 
                       &                                                                             & Assignment 1 & Assignment 2 & Assignment 3 \\ \hline
One                    & 16                                                                          & 70           & 64.49        & 79.53        \\
Two                    & 15                                                                          & 95.41        & 86.25        & 83.75        \\
Three                  & 15                                                                          & 95           & 94.17        & 91.67        \\ \hline
\end{tabular}
\textit{Full score of each assignment is 100.}
\end{table}

One-way multivariate analysis of variance (MANOVA) was applied to examine the differences in student performance across the three groups. Using Pillai's trace, there was a significant effect being detected, V = 0.655, F(6, 84) = 6.823, p $<$ 0.01. The observed statistical power was 0.98.

The MANOVA was followed up with discriminant analysis, which revealed two discriminant functions. The first function explained 98.2\% of the variance, cononical $R^2$ = 0.63, whereas the second explained 1.8\%, canonical $R^2$ = 0.03. In combination, these discriminative functions significantly differentiated Group One from Group Two and Three, $\lambda$ = 0.363, $\chi^2$(6) = 42.513, p $<$ 0.01, but removing the first function indicated that the second function did not significantly differentiate the remaining two groups, $\lambda$ = 0.970, $\chi^2$(2) = 1.280, p $>$ 0.05. In other words, the significant differences detected by MANOVA only existed between Group One and Group Two / Three. No significant difference was found between Group Two and Three.

The findings show that when students only received \textit{What's wrong} feedback, their performance significantly lagged behind their counterparts receiving \textit{Gap} feedback. However, no significant difference was observed between the groups with and without \textit{Hint} feedback.

\subsection{How do automated feedback configurations affect student usage behavior?}

To answer the question "\textit{How do automated feedback configurations affect student usage behavior?}", we conducted a one-question survey at the end of the course to (1) confirm that students indeed used the provided automated feedback, and (2) to learn about group differences in terms of feedback usage. The one question in the survey was:

\begin{displayquote}
How often did you check the feedback on Travis-CI?
\end{displayquote}

A 4-point Likert scale was adopted for the question. Choices for the question include: 

\begin{enumerate}[label=(\alph*)]
  \item Rarely
  \item Sometimes
  \item Often
  \item Always
\end{enumerate}

The four choices corresponded to points ranging from 1 to 4. The average of all 46 students was 3.72. One-way Analysis of variance showed no significant difference across the three groups, F(2, 43) = 0.124, p $>$ 0.05. In other words, students reported that they used the automated formative feedback frequently regardless of which group they were in.

\begin{figure}[H]
\centering
\includegraphics[width=90mm]{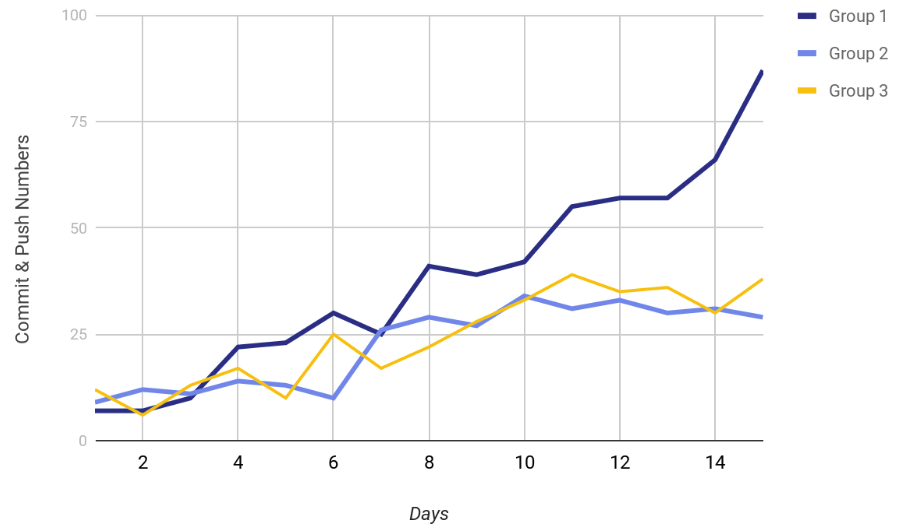}
\caption{An example of commit \& push numbers per group for a programming assignment}
\label{fig:commit}
\end{figure}

To further understand the group difference in feedback usage, we aggregated student feedback seeking behaviors (i.e., commit \& push) by day. When the aggregated commits \& pushes are plotted against the time, there is a clear difference in commit \& push numbers among the three groups. For instance, students had 15 days to work on programming assignment three (see Figure \ref{fig:commit}). In the first five days no clear pattern can be found. In the remaining ten days, Group One committed \& pushed significantly more frequently than Group Two and Three, especially during the last three days ahead of the due date. However, there is no apparent frequency difference between Group Two and Three. Similar effects were observed on all three programming assignments. Overall, students who only received \textit{What's wrong} feedback committed \& pushed more frequently than their counterparts receiving \textit{Gap} feedback. No significant difference was observed between the groups with and without \textit{Hint} feedback.

\section{Discussion}

Among the findings of this study, we would like to highlight two points. First, the observed "system gaming behaviors" among students might be an indicator of ineffective feedback design instead of students abusing automated feedback intentionally. Given that nearly all students reported substantial usage of automated formative feedback, it is safe to assume that when a  student committed \& pushed their code to GitHub, they intended to seek feedback. Based on this assumption, it is obvious that students who only received \textit{What's wrong} feedback sought feedback much more frequently than their counterparts in Group Two and Three. If we only look at the commit \& push numbers of this group of students exclusively, it is tempting to conclude that this group of students gamed the system and abused automated feedback. Many prior studies that had similar observations interpreted it as "intentional system gaming behaviors", and further proposed limiting the number of feedback students can get, or providing no feedback prior to assessment at all \cite{baker2004off, baker2008students, d2006generalizing}. However, when we classified feedback into different types and tested them individually in a controlled study, evidence against this interpretation emerged. Students who only received \textit{What's wrong} feedback sought feedback more frequently, but they did not perform as well as their counterparts receiving more fine-grained feedback. In other words, the reason this group of students sought more feedback is not likely because they were taking advantage of the unlimited feedback. On the other hand, those students might not get the help they needed in problem solving. They kept seeking more feedback simply to use it as a confirmation to see if they successfully passed all test cases, but they were rarely sure if they were on the right track. 

Second, the effectiveness of hints delivered by automated feedback deserves further investigation. Another important finding from the results is that there was no significant difference between students who received and those who did not receive \textit{Hint} feedback in either academic performance or feedback-seeking frequencies. There are different ways to interpret this finding. One possibility is that the \textit{Hint} as a feedback type was not implemented well enough to realize its full potential in this study. Only adaptive hints for top five errors per test case were implemented. For errors outside of this scope no adaptive hints were provided. It is possible that the summarized top five errors were not representative enough to cover the errors students made during this experimental course. As the result, students found hints of little help. Some researchers may argue that the lack of multi-level hints is another reason. We intentionally decided not to implement hints in multiple levels in which the bottom level is closest to revealing direct answers. This design was found to stimulate system gaming behaviors and to be detrimental to student learning in studies on intelligent tutor systems \cite{roll2011improving, baker2008students}. Another possible interpretation is that the comparisons between expected and actual outputs provided sufficient information for students to move forward. It is worth noting that students taking the experimental course were provided multiple channels to have their questions answered, including instructor office hours, teaching assistant office hours, and a dedicated online Question \& Answer platform where students can ask learning questions to both instructors and their peers. The process of debugging, research and fixing errors might take time, but also provide students a valuable opportunities to learn debugging \cite{murphy2008debugging, griffin2016learning}. 

\section{Limitations}

This study is not without limitations. The sample size of this study is comparatively small. Although the sample size was sufficient for a three-group controlled experiment, it is unknown whether the findings can be generalized to CS courses with significantly larger sizes, or outside of the context of CS2. Future studies may consider replicating this experiment in the context of a larger CS course, especially a CS1. Additionally, no qualitative data on how students actually used automated feedback were collected from the experiment. Either in-depth interviews or observations can provide richer information on student usage of automated feedback and the relationship between usage and automated feedback configuration. Future studies are recommended to utilize both quantitative and qualitative methods to answer the research question "\textit{How do automated feedback configurations affect student usage behavior?}".

\section{Conclusions}

This study investigated the essential of meaningful automated feedback for programming assignments using a quasi-controlled experiment. The results revealed that simply knowing what fails does not help students sufficiently, and may stimulate system gaming behavior. Hints were not found impactful on student performance or their usage of automated feedback. In contrast, the gap between the current and expected states seem to provide sufficient information for students to move forward and fix errors in the context of a CS course where multiple support venues were available. We discussed the implications of the findings and further provided guidance on effective automated feedback design based on the findings.

\bibliographystyle{IEEEtran}
\balance
\bibliography{references}

\end{document}